# From Byte to Bench to Bedside: Molecular Dynamics Simulations and Drug Discovery

Mayar Ahmed[1†], Alex M. Maldonado[1†], and Jacob D. Durrant[1*]

[1]Department of Biological Sciences, University of Pittsburgh, Pittsburgh, Pennsylvania, United States of America

[†]These authors contributed equally

*Correspondence: durrantj@pitt.edu

## Abstract
Molecular dynamics (MD) simulations and computer-aided drug design (CADD) have advanced substantially over the past two decades, thanks to continuous computer hardware and software improvements. Given these advancements, MD simulations are poised to become even more powerful tools for investigating the dynamic interactions between potential small-molecule drugs and their target proteins, with significant implications for pharmacological research.

## Introduction

Throughout most of human history, drug discovery relied on trial and error. Modern structural biology has revolutionized the field by enabling rational drug design. This approach uses the molecular structures of disease-implicated targets (typically proteins) to guide the identification and optimization of small-molecule ligands—initial hits that can be further developed into drugs. Structure-based computer-aided drug design (CADD) further augments rational design by using computational methods to drastically reduce the physical experiments required for hit identification, making early-stage drug discovery more cost-effective and efficient.

Traditional CADD techniques have focused on static protein structures. But proteins are highly dynamic in solution, and ligand binding pockets often sample many pharmacologically relevant shapes (i.e., conformations). A given small-molecule ligand may bind to and stabilize only a subset of conformations that complement its shape and specific arrangement of interacting functional groups. Moreover, different ligands often stabilize distinct yet equally valid pocket conformational states. CADD methods that exclusively consider a single pocket conformation thus run the risk of overlooking potential ligands that may bind to alternative conformations.

Atomistic molecular dynamics (MD) simulations have emerged as valuable tools for investigating the conformational diversity of ligand binding pockets. These simulations approximate the complex quantum-mechanical forces that govern atomic motions by representing atoms and bonds as simple spheres connected by virtual springs (1, 2, 3). Researchers have meticulously parameterized the stiffnesses and equilibrium lengths of these springs, as well as the spheres' masses, radii, partial electric charges, etc., so atomic motions evolve realistically under the influence of Newton's laws of motion. The resulting trajectories allow researchers to visualize and analyze protein dynamics at the atomic scale.

CADD researchers routinely use MD simulations to unveil pharmacologically relevant conformational changes, allosteric mechanisms, and binding-pocket dynamics (2, 4, 5, 6, 7). In this comment, we provide a concise overview of the intersection between MD simulations and CADD over the past two decades, emphasizing the advancements that have enhanced our understanding of protein flexibility and its profound impact on drug discovery.

## Generating conformational ensembles

To identify structurally diverse small-molecule ligands that bind to a dynamic binding pocket, CADD must account for multiple physiologically relevant pocket conformations. MD simulations are valuable tools for capturing these continuous conformational changes, including the opening and closing of transient druggable subpockets that are challenging to detect experimentally (8, 9, 10). By clustering the many conformations sampled during an MD simulation (11), one can generate a condensed yet diverse set of representative pocket conformations, known as a conformational ensemble. This ensemble can then be used in subsequent CADD analyses.

## Enriching conformational ensembles by capturing long-timescale dynamics

Longer simulations often reveal more comprehensive conformational ensembles. While shorter simulations are useful in drug discovery, they primarily capture only rapid molecular events such as local fluctuations, surface sidechain rotations, and fast loop reorientations (12). Longer simulations can reveal how slower dynamics such as slow loop reorientations, buried sidechain rotations, and some allosteric transitions (12) impact binding-pocket geometries, revealing druggable conformations absent in shorter simulations.

## Hardware and supercomputing advances

Advancements in computer hardware over the past two decades have enabled much longer simulations. Notably, the adoption of graphics processing units (GPUs) has revolutionized the field by dramatically accelerating calculations (13, 14). Designed initially as highly parallel processors to enhance video game performance, GPUs have been repurposed to accelerate scientific calculations, including those required for faster and more efficient MD simulations.

Ever-increasing supercomputer resources have also enabled longer simulations. Supercomputing power is commonly measured in terms of floating-point operations per second (FLOPS). In 2000, the performance of the world's top 500 supercomputers ranged from 43.8 gigaFLOPS (i.e., billion FLOPS) to 2.4 teraFLOPS (i.e., trillion FLOPS; roughly the performance of an iPhone 14 Pro (15)). By 2010, this range was 24.7 teraFLOPS to 1.8 petaFLOPS (i.e., quadrillion FLOPS), and by 2023, it was 1.9 petaFLOPS to 1.2 exaFLOPS (i.e., quintillion FLOPS) (16).

Thanks to these enabling technologies, researchers can now better explore the pharmacological implications of longer timescale dynamics (17). For example, the first MD simulation in 1977 captured 8.8 picoseconds of bovine pancreatic trypsin inhibitor dynamics (18). It took another 21 years to achieve the first microsecond simulation of a protein in explicit solvent—a remarkable 10-million-fold increase in simulation length (19). In the 2000s, several groups reported explicit-solvent protein simulations on the order of microseconds (20, 21, 22, 23, 24, 25, 26, 27), and since 2010, several millisecond-regime simulations have been reported (28, 29, 30, 31, 32, 33).

Emerging technologies will soon enable even longer simulations. Many computing tasks benefit from application-specific integrated circuits (ASICs), custom-designed chips tailored for a specific task rather than general-purpose use. As the demand for accelerated MD simulations grows among academic and industry researchers, we expect the proliferation of ASICs with optimized architectures explicitly designed for MD acceleration. The remarkable performance gains achieved by the Anton series of supercomputers, which are purpose-built for MD simulations, demonstrate the tremendous performance gains possible through specialized hardware. The latest Anton 3 system achieves a 460-fold speedup compared to general-purpose supercomputers when simulating a 2.2 million atom system (34). Future systems incorporating ASICs, related chips called field-programmable gate arrays (FPGAs), or other specialized hardware (35) could enable routine access to longer, biologically relevant timescales.

**Software advances**

Software advancements have also allowed MD simulations to more thoroughly sample physiologically relevant conformations. These methods bypass the substantial energy barriers that separate different conformational states, rare transitions that typically only occur on longer timescales.

**Rare-event sampling.** Enhanced sampling techniques (36, 37) such as umbrella sampling (37, 38, 39, 40), metadynamics (36, 37, 41, 42), and weighted ensemble path sampling (43, 44, 45, 46) aim to improve sampling along a predefined progress coordinate. This coordinate, also known as a collective variable (37), quantitatively describes the progress towards a target state or set of states (e.g., it provides a quantitative description of the different states and the pathways connecting them). Researchers typically select the progress coordinate based on prior knowledge of the system, often considering factors such as the orientation of or distance between system components.

Other methods enhance sampling without relying on a predefined progress coordinate. These include parallel tempering (also known as replica exchange) (36, 37, 47, 48, 49), simulated tempering (50), simulated annealing (36), integrated tempering sampling (37, 51), and hyperdynamics (also known as accelerated molecular dynamics) (52, 53, 54), among other methods (55). Recently, machine-learning approaches such as autoencoders (56, 57, 58, 59) and other neural network architectures (60) have also gained popularity. These methods map molecular systems onto low-dimensional spaces, where progress coordinates can better capture complex rare events (61). Though the specific details of each method vary and are beyond the scope of this comment, they all share the goal of algorithmically improving sampling efficiency.

**Machine-learning structure prediction.** To enhance binding-pocket sampling, some have coupled MD and AlphaFold (62), a recently developed machine-learning approach for protein structure prediction. MD is critical because AlphaFold and related methods (62, 63) often struggle to position sidechains accurately enough for effective CADD (64, 65, 66), but brief MD simulations can correct misplaced sidechains. These simulations, which often involve placing a crystallographic ligand within the pocket to encourage transitions to the *holo* conformation (67), can substantially improve the accuracy of subsequent ligand-binding predictions (67, 68).

Modified AlphaFold pipelines also overcome the default implementation's tendency to converge on a single conformation, making it possible to predict entire conformational ensembles (69). One approach is to use only random subsets of all available homologous sequences during the multiple sequence alignment step or to limit the structural templates the AlphaFold model considers (70, 71). The goal is to enable broader sampling by reducing the evolutionary signal, thus increasing uncertainty. Other strategies include masking columns in the alignment (72), running AlphaFold separately on clusters of similar (non-random) sequences (73), or providing structural templates that include related proteins in the desired conformational state (74, 75). The diverse conformations these approaches predict can then serve as seeds for short simulations (76), bypassing the need for long-timescale simulations that would otherwise be required to transition between the conformational states.

**Machine-learning force fields.** Simulations that capture quantum effects have also benefited from recent software acceleration. Classical MD simulations represent atoms and bonds as simple spheres connected by virtual springs (1, 2, 3). This classical approach imposes parameterized analytical approximations that overlook crucial interactions such as electron correlation, nuclear quantum effects, and electron delocalization. Consequently, classical simulations cannot model chemical reactions, nor can they account for some subtle non-covalent effects that may impact ligand binding. Accurately accounting for these factors requires computationally intensive quantum mechanical (QM) methods such as Kohn-Sham density functional theory (DFT).

Machine-learning approaches have the potential to accelerate these calculations without imposing analytical constraints. Given enough conformationally diverse and sufficiently accurate high-level QM calculations for training, these approaches can effectively learn arbitrary potential energy surfaces. General-purpose machine-learning force fields trained on millions of small-molecule DFT calculations, such as ANI-2x (77), can be used for biomolecular simulations. Although these force fields enable otherwise intractable QM calculations, they are still much slower than classical methods. We expect ongoing advancements in computational speed and accuracy to broaden their adoption in the future.

### Enriching conformational ensembles via mesoscale simulations

Simulating protein targets embedded in larger macromolecular complexes or realistic subcellular environments can also help identify more complete ensembles of physiologically relevant pocket conformations. These simulations better account for the impact that interactions with macromolecular partners have on binding-pocket geometries. Similarly, crowding effects (78, 79, 80) can significantly impact protein dynamics.

Over the past two decades, researchers have made remarkable progress in simulating increasingly larger systems (81). The first atomistic MD simulation performed in 1977 had fewer than 1,000 atoms (18). By 2002, the accessible system size had increased over 100-fold (e.g., the aquaporin channel, ~100,000 atoms (82, 83)). This size increased another 10-fold by 2006 (complete satellite tobacco mosaic virus, ~1 million atoms) (84), and yet another 10-fold by 2017 (HIV-1 capsid, ~64 million atoms) (85). In recent years, several simulations have surpassed the 100 million atom mark (photosynthetic chromatophore, ~136 million atoms (86); H1N1 influenza virion coat, ~160 million atoms (87); SARS-CoV-2 viral envelope, 305 million atoms (88, 89)). Notably, two groups have reported billion-atom simulations (the entire GATA4 gene locus (90); a SARS-CoV-2 virus within a respiratory aerosol (89)). These advancements in simulating larger systems have provided unprecedented opportunities to explore complex biological phenomena at an atomic level.

# Ligand pose prediction

Molecular docking programs are valuable CADD tools that predict the binding mode or "pose" of a small-molecule ligand within a specific binding-pocket conformation (91, 92, 93). Traditionally, pose prediction has relied on a single static protein structure derived from techniques such as X-ray crystallography. Although single-conformation docking can effectively identify true ligands, it fails to account for the possibility of alternative but equally valid pocket conformations. Moreover, many proteins lack co-crystallized ligands, making it challenging to experimentally determine their ligand-amenable *holo* states for use in CADD. In some cases, pharmacologically relevant cryptic pockets are entirely collapsed unless bound to a ligand (8, 9, 10). These limitations of single-conformation docking underscore the importance of methods that can better account for full protein dynamics.

In recent years, molecular docking studies have increasingly incorporated ensembles of diverse binding pocket conformations, often sourced from clustered MD simulations (11, 91, 94). Ensemble docking, also known as relaxed-complex-scheme docking (95, 96, 97), produces a spectrum of scores for each compound by docking each into multiple structures rather than just one. One then converts this spectrum into a single score, such as the ensemble-average or ensemble-best score, which is used to rank and prioritize compounds for subsequent experimental testing.

MD simulations have also emerged as valuable tools for validating already docked poses (98). Though docking is widely used for drug discovery, the accuracy of docked poses is sometimes lacking. To address this shortcoming, researchers perform brief MD simulations of the predicted protein/ligand complex and monitor the ligand's drift from its initial position (99, 100, 101, 102, 103, 104, 105). Correctly posed ligands tend to have greater stability, but incorrectly posed ligands often drift within the binding pocket.

## Predicting binding free energies

MD simulations also play a crucial role in predicting ligand binding free energies. Early-stage drug discovery aims not only to discover how small-molecule ligands interact with target proteins but also to assess their binding strength. Simulations provide valuable insights that help prioritize the most promising compounds for subsequent experimental validation and optimization. Here, we focus on MM/GB(PB)SA and alchemical methods, though many other notable approaches have also been developed (98, 106).

### MM/GB(PB)SA

MM/PBSA and MM/GBSA (107, 108) are popular methods for calculating binding free energies. These methods rely on one or more simulated trajectories of protein, ligand, and protein/ligand complexes. Selected frames from these simulations are used to calculate average binding-induced changes in the molecular mechanics energy and the solvation energy, per Poisson-Boltzmann or generalized-Born surface-area calculations. Optionally, one can also account for conformational entropy, often via normal-mode analysis (109).

Although the MM/GB(PB)SA methods were developed over twenty years ago (110, 111), recent advancements in machine learning have improved their accuracy and efficiency. Machine learning has been used to intelligently select simulation frames for subsequent MM/GB(PB)SA calculations (112). Additionally, it has been used to refine the calculation of MM/GBSA energy terms by assigning optimized atom-specific dielectric constants (113) or altering how the individual electrostatic, van der Waals, and solvation terms are combined into final free-energy estimates (114). On a more global level, machine learning has also been used to help researchers optimally select the number of MD and MM-PBSA runs to best strike a balance between accuracy and the need to conserve computational resources (115).

### Alchemical simulations

Though MM/GB(PB)SA methods are widely used, they consider only the bound and unbound states, neglecting the influence of intermediate states on binding energy. Alchemical simulation methods such as thermodynamic integration (TI) (20) and free energy perturbation (FEP) (116) can overcome this challenge, albeit at a higher computational cost. These methods calculate relative free energies by gradually eliminating the nonbonded interactions between the ligand and environment (e.g., protein), effectively disappearing the ligand atoms during simulation (117). By observing the system's response to these nonphysical changes, one can estimate the relative binding free energies (118, 119, 120).

In recent years, advancements driven by machine learning have accelerated free-energy predictions through alchemical simulation. Notably, recent work has shown that AlphaFold-predicted protein models are often accurate enough to support FEP calculations (121), greatly expanding the drug targets that can be studied using this technique. Additionally, active learning has proven instrumental in reducing the number of alchemical calculations necessary to screen large chemical libraries in search of novel small-molecule inhibitors. This approach has successfully identified inhibitors of various targets, including the SARS-CoV-2 papain-like protease (122, 123), TYK2 kinase (124), Wee1 (123), and cyclin-dependent kinase 2 (123, 125). Cloud-computing workflows tailored specifically for alchemical calculations have further enabled studies such as these (126, 127).

## Conclusion

In recent decades, MD simulations and structure-based CADD have benefited from remarkable advancements in computational power, software development, and machine learning. Using these techniques, researchers can better capture the dynamics of ligand binding pockets by simulating ever larger systems for ever longer timescales. The conformational ensembles these simulations reveal enable accurate ligand docking and binding-affinity prediction. As specialized hardware and algorithmic developments become increasingly accessible, the impact of MD simulations on early-stage drug discovery is poised to grow even further.

## Declarations

### Acknowledgments

We acknowledge assistive writing technologies such as Grammarly, OpenAI's ChatGPT, and Anthropic's Claude, which we used to refine text during manuscript preparation. The listed authors thoroughly reviewed, revised, and selectively implemented any suggested edits to ensure accuracy, consistency, and clarity. The responsibility for the paper's content and quality remains with the authors alone.

### Funding

This work was supported by the National Institute of Health (R01GM132353). The content is solely the responsibility of the authors and does not necessarily represent the official views of the National Institutes of Health. The funders had no role in the study design, data collection and analysis, decision to publish, or preparation of the manuscript.

### Availability of data and materials

Not applicable.

**Competing interests**

The authors declare that they have no competing interests.

**Author contributions**

M.A. performed background research, including an extensive literature search, and provided input into the manuscript's organization. A.M.M. contributed to the text describing enhanced sampling techniques, machine-learning force fields, and alchemical methods. J.D.D. drafted and edited most of the text. All authors have read and agreed to the content.